# The Tyranny of Possibilities in the Design of Task-Oriented LLM Systems: A Scoping Survey

Dhruv Dhamani
*University of North Carolina, Charlotte*
ddhamani@charlotte.edu

Mary Lou Maher
*University of North Carolina, Charlotte*
m.maher@charlotte.edu

*Abstract*— This scoping survey focuses on our current understanding of the design space for task-oriented LLM systems and elaborates on definitions and relationships among the available design parameters. The paper begins by defining a minimal task-oriented LLM system and exploring the design space of such systems through a thought experiment contemplating the performance of diverse LLM system configurations (involving single LLMs, single LLM-based agents, and multiple LLM-based agent systems) on a complex software development task and hypothesizes the results. We discuss a pattern in our results and formulate them into three conjectures. While these conjectures may be partly based on faulty assumptions, they provide a starting point for future research. The paper then surveys a select few design parameters: covering and organizing research in LLM augmentation, prompting techniques, and uncertainty estimation, and discussing their significance. The paper notes the lack of focus on computational and energy efficiency in evaluating research in these areas. Our survey findings provide a basis for developing the concept of linear and non-linear contexts, which we define and use to enable an agent-centric projection of prompting techniques providing a lens through which prompting techniques can be viewed as multi-agent systems. The paper discusses the implications of this lens, for the cross-pollination of research between LLM prompting and LLM-based multi-agent systems; and also, for the generation of synthetic training data based on existing prompting techniques in research. In all, the scoping survey presents seven conjectures that can help guide future research efforts.

## I. Introduction

*Large Language Models* (LLMs) are a recent development in Generative Artificial Intelligence that can mimic human-like behavior [1], especially in conversations [2]. LLMs have also exhibited a kind of general intelligence [3], [4]. Central to harnessing the capabilities of LLMs is the concept of *prompting*, a strategy that significantly influences task performance by instructing LLMs in specific ways [5]. For example, it is found that asking LLMs to share their thoughts step-by-step while attempting a task improves performance [6], [7].

Equally important is *augmentation*, where messages to LLMs are augmented with useful knowledge (retrieval) [8], examples of solved problems (in-context learning) [9], and know-how for tool-use [10]. These tools are generally computational procedures that influence an environment (either digital or real) or can offload computations that LLMs are unsuitable for, such as math [11]. LLM's primary involvement here is to choose which procedure (tool) to use and what parameters to pass to the procedure. Well-designed tools can provide agency to LLMs - the capability to influence their environment [12]. The resulting LLM-based agent can be instructed to solve useful tasks, resulting in a task-oriented LLM system[1].

Any non-trivial task can be broken down into sub-tasks, meaning any development of task-oriented LLM systems will lead to the development of systems where multiple such systems collaborate on solving sub-tasks oriented towards solving a root task. The sub-systems of a task-oriented LLM system are also meant to solve certain tasks, and indeed, LLM systems have been explored where the augmented knowledge, examples, and tools are generated by other LLMs, LLM-based agents, or through LLM involvement. Even breaking down a task into sub-tasks is a task in itself, meaning the possibilities are endless in the design of task-oriented LLM systems.

---

[1] In Xi *et al.* [13], the authors describe *task-oriented deploments* of LLM-based agents, which we generalize to simply task-oriented LLM systems.

In the face of such layered but infinite complexity, it is important to be able to predict uncertainty in LLM task performance (so complexity can be layered on the fly as needed) and create practical metrics to evaluate task-oriented LLM systems. Thus, this scoping survey aims to assess the current state of research in this area and identify the most pressing research questions and gaps.

The current draft of this survey covers the current state of research in Large Language Model Augmentation (see § III.A), Prompting (see § III.B), and Uncertainty Estimation (see § III.C) - with a focus on how these affect the design of task-oriented LLM systems.

The survey is non-exhaustive, attempting to re-organize and summarize select research. The primary database for this survey is arXiv, as this is a rapidly evolving area of research with most relevant research published as pre-prints within the last year.

The paper is organized into the following sections -

1. **Exploring the Design Space** (Section II): This section explores the design space of task-oriented systems through a thought experiment focusing on a complex software development task to test the limits of current LLMs. It includes a definition of a minimal task-oriented LLM system, an analysis of design parameters, task description and assumptions, different LLM system configurations, and a discussion of their hypothetical effectiveness.

2. **Current Research for Select Design Parameters** (Section III): We share the current state of research in three key areas: LLM Augmentation, Prompting, and Uncertainty Estimation; and examine their potential impact on the design of task-oriented LLM systems.

3. **Discussion** (Section IV): The discussion section interprets the implications of our findings for future research in task-oriented LLM systems. It covers the evaluation of LLM systems, differentiating and defining Linear and Non-Linear Contexts in LLM systems, the concept of Agent-Centric Projection of Prompting Techniques, and its potential for Synthetic Training Data Generation.

4. **Conclusion** (Section V): The final section summarizes the paper, presents its limitations, highlights key insights and implications for future research in the field of task-oriented LLM systems.

## II. Exploring the Design Space

In this section, we aim to explore the design space of task-oriented LLM systems through a thought experiment involving a task that is beyond the capabilities of current LLMs, such as developing a large, complex software project based on a given set of project requirements. Software development is an iterative process. As issues, or opportunities to refactor and improve, surface, we revisit prior work and make the required changes. Considering this, it is unlikely any complex software project can be completed and shared within a single response by an LLM.

This section begins by defining a **minimal task-oriented LLM system** (§ II.A). We then explore various **design parameters** (§ II.B) that affect the task performance of such a system. This is followed by a detailed **task description** and our underlying **assumptions** (§ II.C). We then carry out the **thought experiment** (§ II.D) by hypothesizing about the effectiveness of various system configurations in executing the task. Finally, we will **discuss** (§ II.E) the outcomes of our hypothetical scenarios and their broader implications.

### II.A. *Minimal Task-oriented LLM System*

Before we proceed, we need to first define a minimal task-oriented LLM system. A minimal task-oriented LLM system is a minimal LLM system that is instructed to solve a task. Thus, we will be defining a minimal LLM system.

Large Language Models are autoregressive models that accept input tokens and use them as history (often referred to as context), to compute probabilities of all tokens in their vocabulary as the next token. We can sample from this probability distribution using a sampling/decoding procedure, to generate text. This process is then repeated until the LLM predicts a special token, or a special sequence of tokens, that indicates the end of the text [14][2].

We call this a *barebones LLM system* (see Figure 1) as it only contains the minimal components needed for text generation, with no additional components to help with context management. Every time an LLM is prompted with context $C_n$, it generates a response $R_n$ which would need to be stored in the context $C_{n+1}$ for the next prompt, assuming multiple rounds of instruction and response generation are required.

---

[2]The *generative AI system* description in S. Feuerriegel *et al.* [14] includes any UI components as part of the generative AI system, we use a modified definition that only includes the language model and sampling/decoding procedure here.

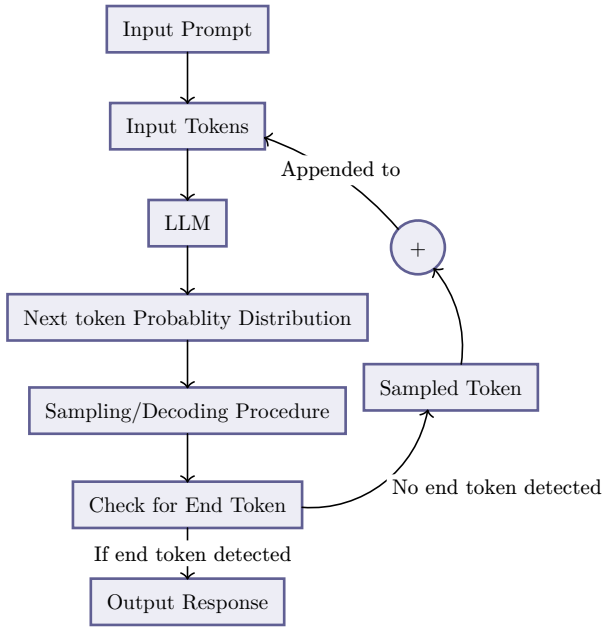

Figure 1: A barebones LLM system.

For systems oriented towards solving even moderately complex tasks, context management becomes quickly cumbersome. For some LLM systems, a transformation is applied on response $R_n$ before it is stored in context $C_{n+1}$ as described in Section III.A.2. In S. Saha *et al.* [15], the authors describe a system where $m$ branches are created from a LLM response to a prompt $C_n$, producing a set of $m$ responses $r = \left\{R_{n_1}, R_{n_2}, ..., R_{n_m}\right\}$. Then, they take all of $r$, and transform it to a prompt $C_{n+1}$, in which they instruct the LLM to merge all responses in $r$ into a single response $R_{n+1}$, which potentially needs be stored in context $C_{n+2}$ for the next prompt. A similarly complex system is described in X. Ning *et al.* [16], and even more examples are shared in Section III.B.

If we define a minimal LLM system without describing how context is managed, it would then be too difficult to compare different systems and apply learnings from one researched system to another. Because of this, we include a description of a minimal context management sub-system within our definition of a minimal task-oriented LLM system.

Specifically, we include a *context store* CS. Initially, the context store CS is empty, and the first time an LLM is provided with context $C_1$ to generate $R_1$, both prompt and response are permanently appended to CS. For all future requests, the LLM is first provided with a sliding window of content from CS as context, to which it appends the prompt $C_n$ to generate $R_n$. Once response $R_n$ is generated, both prompt and response are again permanently appended to CS. The model then closely matches messaging, where the context store CS is the chat history, the users of the LLM *send* messages, and the LLM *responds* to messages.

This description of a context store is similar to the latest Assistants API from OpenAI [17], and to the most frequent use-case for LLMs currently - as a kind of helpful chatbot.

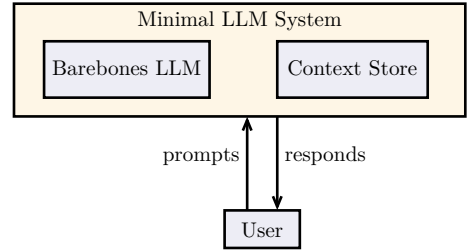

Figure 2: A minimal LLM system that includes a context store.

We discuss the implications of this composition of a minimal LLM system in § IV.C.

## II.B. *Design Parameters*

The following design parameters may be tweaked to influence performance in task-oriented LLM systems, based on current research –

1. **Training Data**: LLMs show degraded performance in tasks that are underrepresented in their training data (out-of-distribution tasks) [18]. While more capable LLMs can use tools without specialized training [19] and supervised fine-tuning can improve performance [20]; it is unclear if tasks, where *extensive tool-use* is required, are well represented in their training data, enough to allow them to be solved by LLMs [21]. *Extensive tool-use* refers to the capacity to predict outcomes in the environment over many instances of planned tool-use, where the influence of the tool may be asynchronous and out-of-order.

    *Theory of Mind* refers to the capacity to ascribe and track mental, unobservable states of other people. Some evidence has been presented that shows LLMs may have partially gained this capacity through their training [22], [23], but it is unclear if this capacity is sufficient for tasks that involve extensive collaboration. *Extensive collaboration* refers to a large number of LLMs-based agents collaborating, in a manner that may require some agents to ascribe and track unobservable states of multiple other agents.

    Finally - task decomposition, planning, reasoning, reflection, and collaboration, are all examples of skills that may affect task performance in task-oriented LLM systems. An exact list of such skills,

their importance, minimum requirements, and how to encourage such skills during training are unclear.

2. **Alignment Techniques**: Most LLMs used today are aligned to follow instructions and respond according to human preferences [24], and also for two-party conversations [25]; it is unclear if this is ideal, or if alternative alignments are required for LLM use in task-oriented LLM systems - where multiple LLMs (or LLM agents) may collaborate to solve tasks (so an $N$-party conversation), and there may be minimal human involvement.

3. **Context size and context utilization**: While highly capable LLMs can accept longer context lengths of up to 128-200 thousand tokes [26], [27]; studies have shown that LLMs may not actually use this context well [28]. It is unclear if and how this would affect long-running task-oriented LLM systems where large context sizes may be required. We will expand on why this may be important during our thought experiment (§ II.D).

4. **Large Language Model Augmentation**: Large Language Model Augmentation refers to how LLMs are augmented with knowledge, examples, and tools. Each can significantly affect task performance [8], [9], [19]. An elaborate discussion follows in Section III.A.

5. **Agency and Autonomy**: LLMs can be encapsulated in systems and placed in environments such that they have agency, i.e., they can influence their environment. Such systems are called LLM-based agents, and multiple such agents can collaborate forming multi-agent systems [13], [29], [30]. What autonomy paradigms (event-loops), and collaboration paradigms (between agents) are suitable for which task-oriented LLM systems is unclear.

6. **Prompting**: Prompting refers to how LLMs are instructed to solve tasks. Prompting has been shown to significantly affect task performance [31], [32]. An elaborate discussion follows in Section III.B.

7. **Sampling/decoding procedures**: LLMs generate text by sampling from a probability distribution over all tokens in their vocabulary. Suboptimal sampling strategies may compound when generating large amounts of text, leading to catastrophic failure or leading the LLM to stray from its original instructions (there is limited support for this in the domain of long-form storytelling [33]). It is unclear if this is a significant issue for long-running task-oriented LLM systems.

For our thought experiment, we will consider various configurations of a task-oriented LLM system and hypothesize the effectiveness of each configuration at the task. To keep the complexity and length of the thought experiment to a manageable level, we will only vary the LLM used, agency, autonomy, collaboration, and tool augmentation between the configurations we discuss.

### II.C. Task Description and Assumptions

1) **Task Description**: A human shares a list of project requirements for a complex software project as a prompt, instructing the task-oriented LLM system to develop the described project. The system is to start and keep working until (a) it "decides" it is done and shares the final deliverable (b) it "decides" the task is outside its capabilities and gives up. We assume the final response from the LLM system (without access to tools), unless it has decided to give up, would be a list of file names and paths, and contents of the file. For tool-augmented LLMs, we assume the final response would be a list of files and paths, and that the system would have written the contents of the files to disk. Success or failure in the task will be determined by whether the deliverable meets the project requirements. The LLM will not be augmented with any tools unless specified otherwise.

2) **Hypothesized Effectiveness**: As success at this task is boolean, it will not be easy to differentiate between all the failed configurations and all the successful configurations. Because of this, we will use a different metric – hypothesized effectiveness. We define it as the hypothesized success rate when solving 100 such software development tasks, where each task is iteratively more complex. The simplest task is simple enough to be completed in a single GPT-4 response, and the most complex task is as described above.

3) **Assumptions**: The LLM system is expected to continually write new code and periodically revisit and edit old code (as software development is an iterative process). Because of this, for this thought experiment, we assume that for the more complex tasks, the LLM system would have to generate at least 2 million tokens in total. Also, there is the possibility that an important interface or API may be defined in a file at the beginning of the context, and may be needed towards the end. Because of this, we also assume that ideally, the entirety of this 2 million token-long context is relevant for the successful generation of future tokens. This assumption does not necessarily apply to tool-augmented LLM systems, as we will discuss in § II.D.

For this thought experiment, we are also in an LLM that surpasses OpenAI's GPT-4, which we will call GPT-6. We make the following assumptions about GPT-6, to create a scenario where we have access to orders of magnitude more compute but haven't managed to make any meaningful progress in unlocking new emergent capabilities in LLMs (so no significant advances in training, alignment, etc.) -

1. GPT-6 is a text-only LLM, for ease of comparison.
2. GPT-6 is based on minor variations of the same fundamental architecture, with most improvements in capabilities coming from orders of magnitude higher compute.
3. Because of this, GPT-6 can operate decently well over a context length of 2 million tokens.
4. GPT-6 is a better reasoner and planner, knows more facts, improves on theory of mind tasks, and has a better world model, but does not consistently outperform expert humans.
5. GPT-6 still does poorly on out-of-distribution tasks (tasks it hasn't seen examples of during training) and hasn't shown any dramatic emergence of new capabilities. Because of this, we still assume it takes around 2 million tokens to complete the more complex tasks task and is not suddenly capable of completing it in one response.
6. GPT-6 does not have self-awareness of its limitations, and because of this, we will assume it will not creatively use any available tools to completely bypass its limitations.

***Superintelligent LLM:*** Finally, we would need a superintelligent LLM. We assume superintelligence to be "a kind of intelligence which is able to understand, to know and to predict everything. There are no boundaries and no limits for its thinking, creativity, and innovation.", as described in K. Szocik *et al.* [34].

## II.D. Thought Experiment

Consider the following scenarios, where various configurations of GPT-4, GPT-6, or a Superintelligent LLM are being used to complete this task and let us try to rank them in the order of hypothesized effectiveness at this task.

1) ***Single LLMs:***

1.a) ***Scenario S1 (Superintelligent LLM):*** Superintelligent LLM was assumed to be one that can correctly predict complex outcomes, and because of this, it should be able to complete this task within a single response. Even though software development is an iterative process, the superintelligent LLM should be able to correctly predict outcomes resulting from the code it is writing, as it is being written, and make any changes necessary. As a result, it should have perfect hypothesized effectiveness at this task.

1.b) ***Scenario S2 (Lone GPT-4):*** OpenAI's GPT-4 is being instructed to complete the entire task on its own. As it does not have access to tools, it would not be able to execute the code it is writing and check if it is working as intended. In the absence of this capability, GPT-4 would have to resort to predicting outcomes, and it is extremely unlikely that GPT-4 would manage to complete all but the simplest of tasks.

1.c) ***Scenario S3 (Lone GPT-6):*** We assumed GPT-6 can work well with context lengths of 2 million tokens, which is how long we assume the most complex task takes to finish. However, without access to any tools that would allow it to execute its code/tests, it won't be able to check its work and would have to rely on predicting whether the code works.

Being able to correctly predict complex outcomes can be considered an element of superintelligence. We assumed GPT-6 does not approach superintelligence, and thus there would be at least one or more failures to correctly predict outcomes which lead to incorrect assumptions in future work, eventually leading to failure. GPT-6 would likely incorrectly predict it succeeded too, in many cases.

We would have to assume that GPT-6 fails frequently, and thus the hypothesized effectiveness is low.

2) ***Single LLM-based Agent:***

An autonomous agent is a system situated within and a part of an environment that senses that environment and acts on it, over time, in pursuit of its own agenda and so as to effect what it senses in the future.
— S. Franklin and A. Graesser [35]

2.a) ***Scenario A1 (Lone Autonomous GPT-4 Agent):*** GPT-4 is augmented with tools such that it can execute code, write unit tests, run them, observe the output, and use it as feedback for future iterations. This configuration of GPT-4 also has the autonomy to keep working until it decides to stop.

The concept of autonomous large language model-based agents has garnered significant interest [30]. However, the success of current implementations is somewhat limited. For instance, in H. Yang *et al.* [36], during a simulated *WebShop* task requiring an autonomous agent to make accurate purchase decisions in an online store according to user specifications, GPT-4 achieved success in less than half of the attempts. The reason for this might be that while GPT-4 is capable of individual instances of tool-use based on given instructions during conversations (where each message from the user can be thought of as a new instruction) — it seems to struggle with *extensive tool-use*.[3]

GPT-4 would also likely be limited by its limited context length here, which would prove inadequate over the assumed 2 million tokens it would take to finish the most complex task. It is unlikely we achieve success frequently in this scenario, though the hypothesized effectiveness is likely better than Scenario S2.

2.b) ***Scenario A2 (Lone Autonomous GPT-6 based Agent):*** Given our primary concern for failure in the previous Scenario S3 was that GPT-6 would not have access to tools and would fail to correctly predict outcomes - if you give GPT-6 access to these tools, then we must assume that likely it succeeds more frequently.

However, something else to consider is the challenges in generating a large number of tokens (we assumed 2 million tokens). The output of an LLM is a probability distribution over all tokens in its vocabulary, and a token needs to be sampled or chosen from this distribution as the predicted next token [14]. Suboptimal choices made during sampling may compound when generating $\geq 2$ million tokens, leading to catastrophic failure or leading the LLM to stray from its original instructions. Perhaps creating a functioning software project, but not one that objectively meets the given project requirements.

At the same time, since GPT-6 has access to tools, where it can read, write, and run code – it may be that it never needs to generate so much text. For example, when editing previously written files, instead of generating the new contents, it could output a diff and apply it. Perhaps a significant number of these tokens would be output from tests being run, stack traces, logs - tokens it didn't itself generate and just used as context (meaning the risk from suboptimal sampling is mitigated as it needs to sample fewer tokens).

So, it seems in this scenario, success on the more complex tasks is dependent on how well GPT-6 can utilize its long context while generating large amounts of text, or it depends on how good GPT-6 is at planning and making optimal use of its tools. Examples of extensive tool-use[3] are likely not prevalent in pre-training or fine-tuning data (at least in current LLMs), so it may be fair to treat extensive tool-use as an out-of-distribution (OOD) task. We assumed that GPT-6 had no new emergent abilities, so we will assume it still lacks capabilities to carry out extensive tool-use.

All in all, the lone GPT-6 agent (based on our assumptions) should fare better than the lone GPT-4 agent but still fail as the tasks get more complex.

3) ***Multiple LLM-based Agents:***

3.a) ***Scenario M1 (Multiple GPT-4 based Agents):*** It is found that when relatively complex tasks are broken down into simpler subtasks, LLMs are better at completing the complex root task [6], [7]. Because of this, there is a lot of interest in having complex tasks lone LLMs cannot solve on their own be solved by multiple LLM-based agents that collaborate [13], [37], [38].

We had assumed it would take over 2 million tokens to complete some of the complex tasks, because of the iterative nature of software development. In the previous scenarios, all, or most of the 2 million tokens had to be generated by the same LLM.

On the other hand, if it takes $2.5 - 3$ million tokens for multiple LLM-based agents to complete the same task successfully (with additional collaboration overhead), these 3 million tokens would be divided among $N$ agents, meaning on average each agent would need to handle $\frac{3}{N}$ million tokens. And each time one of these agents solves a task, they're likely solving a simpler task that requires simple and more obvious use of tools.

Because of this, we hypothesize that multiple GPT-4 based agents collaborating are well-suited to this task, perhaps matching the previous lone GPT-6 agent scenario (§ II.D.2.2), as every agent is solving simpler sub-tasks LLMs are more suitable for.

However, complex software development projects would likely still be too difficult for currently available multi-agent frameworks. It could be that the number of agents $N$ that can effectively collaborate is not sufficient to bring the final $\frac{3}{N}$ million tokens that each agent needs to work with low enough.

As we make progress in task decomposition and multi-agent collaboration, we will be able to increase the number of agents $N$ that can effectively work together. We hypothesize that as $N$ increases, the success rate at this task also increases.

3.b) ***Scenario M2 (Multiple GPT-6 based Agents):*** Multiple GPT-6 based agents collaborating should be able to complete most tasks successfully, as it would only take a few agents to bring the final $\frac{3}{N}$ million tokens that each agent needs to work with, due to the higher effective context length of GPT-6. Each agent would have to generate fewer tokens and make simpler decisions as each would have the responsibility to solve a simpler subtask (so lower requirements for the capacity of *extensive tool-use*). As the number of agents $N$ does not need to be high, *extensive collaboration* capacity of GPT-6 would also not be as important. All this should allow them to complete most, if not all, tasks successfully.

---

[3]Tasks that require the capacity to predict outcomes in the environment over many instances of planned tool-use, where the influence of the tool *may* be asynchronous and out-of-order.

*II.E. Discussion*

In this section, we consider the implications of our thought experiment in more depth. The thought experiment provided a hypothetical yet systematic exploration of various configurations of a task-oriented LLM system. By evaluating and ranking the effectiveness of each configuration, we can derive a clearer understanding of their potential. This analysis not only helps in visualizing their comparative strengths and weaknesses but also forms the basis for the three conjectures we propose, aimed at guiding future research in task-oriented LLM systems. Beginning with the configurations, we arrange them in descending order of their hypothesized effectiveness as follows –

1. Scenario S1 – Superintelligent LLM (§ II.D.1.1)
2. Scenario M2 – Multiple GPT-6 based agents (§ II.D.3.2)
3. Scenario M1 – Multiple GPT-4 based agents (with high value of $N$)[4] (§ II.D.3.1)
4. Scenario M1 & A2 – Multiple GPT-4 based agents (with low value of $N$)[4] and Lone Autonomous GPT-6 based agent (§ II.D.3.1 and § II.D.2.2) (**tied**)
5. Scenario A1 – Lone Autonomous GPT-4 Agent (§ II.D.2.1)
6. Scenario S3 – Lone GPT-6 (§ II.D.1.3)
7. Scenario S2 – Lone GPT-4 (§ II.D.1.2)

Here, Scenario S3 is ranked below Scenario A1 because we assumed without access to tools to run the code being written, the Lone GPT-6 would often incorrectly end up assuming it has met project requirements while its code might not even compile. In contrast, in Scenario S3, while it may not succeed more often, it would more frequently and "correctly" give up when it realizes through its tools that it has failed.

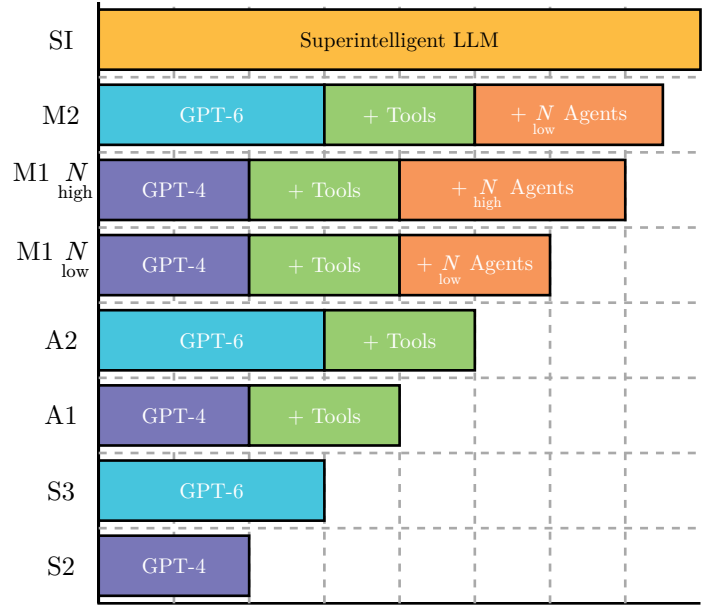

Figure 3: Visualization of hypothesized effectiveness of design scenarios in our thought experiment.

Based on our hypothesis that multi-agent configurations outperform the single LLM or LLM-based agent configurations, we have developed the following conjectures.

> **Conjecture 1**
>
> Autonomous, multi-agent collaboration allows less capable tool-augmented language models to surpass more capable tool-augmented language models, as the number of collaborating agents increases; given these less capable tool-augmented LMs have a threshold level of capablities.

Here, "less capable" and "more capable" refer to our shared understanding of LLM capabilities through benchmark evaluations, such as the ones described in Z. Guo *et al.* [39]. The threshold level of capabilities or skills required here would likely relate to reasoning, planning, task-decomposition, extensive tool-use and collaboration, etc.

However, we have already explored augmenting capable LLMs such as GPT-4 with tools [40] and multi-agent collaboration [38], [37][5]. While we do find that multi-agent systems generally outperform single-agent systems, the magnitude of improvement isn't as significant as we would expect from our thought experiment. Thus we develop 2 more conjectures.

---

[4]where $N$ is the number of agents that can effectively collaborate

[5]S. Hong *et al.* [37] details and evaluates "MetaGPT", a multi-agent collaborative framework on this exact task of software development (though with differing level of complexity), comparing it with a lone GPT-4 system.

> **Conjecture 2**
>
> Multiple LLM-based agents working together should be more capable than current research suggests, and their relative lack of success warrants investigation.

> **Conjecture 3**
>
> Even if we never discover an architecture better than current LLMs, or better training algorithms, or it turns out that scaling up LLMs and their training data does not lead to any new emergent abilities; we can still be able to achieve useful autonomous AI agents through -
> (a) larger context sizes and better context utilization.
> (b) ensuring extensive collaboration between LLM agents and extensive tool-use is "in-distribution", i.e., well represented in its training data.
> (c) sampling/decoding strategies that work well for large context lengths.

## III. Current Research for Select Design Parameters

The current draft of this survey focuses on the current state of research in **LLM augmentation**, **prompting**, and **uncertainty estimation**. While there wasn't an explicit focus on these three areas during our motivating thought experiment, we include these because of the following observations about their relevance to the successful performance of task-oriented LLM systems.

- **LLM augmentation** allows LLMs to use tools, and additional knowledge to help guide tool-use decisions (as we describe in § III.A). Tool-use was a key component of our thought experiment, and we found that it was a significant factor in determining the success of LLMs at the task.
- **Prompting** is how LLMs are instructed (as we describe in § III.B). These instructions may pertain to how to solve tasks, what tasks to solve, how to collaborate, etc. Prompting is a key component of any task-oriented LLM system.
- **Uncertainty estimation** is how LLMs may estimate the certainty of success in a task (as we describe in § III.C). This may be key in multi-agent systems, where ideally each agent solves a simpler sub-task. An early indication of potential failure on sub-tasks may help the "multi-agent team" adapt or the individual agent ask for help. It may also help the system decide when to give up correctly.

### III.A. *LLM Augmentation*

We define ***LLM Augmentation*** as the act of adding additional knowledge/information to the LLM's input context (or prompt), or the act of adding additional knowledge/information to the LLM's output context (or response).

1) ***LLM Input Augmentation:***

LLMs are autoregressive models, and they cannot always generate the *correct* token, only the token their training data suggests is quite likely. Because of this, they are prone to hallucinations, which can make many applications of LLMs impractical [41].

One way to mitigate this is Retrieval Augmented Generation (RAG), where the LLM's input context is augmented with information retrieved from a knowledge base that grounds the LLM's response [42] in factually correct knowledge. It is important to note that there are many different approaches to LLM Input Augmentation.

1) **Constant Augmentation**: Here, we always pass the same additional knowledge or information to the input context, regardless of its context. This can be used for adding memory, adding examples, and providing knowledge about tool use.
   a) **Adding Memory:** This is most frequently in the form of *System Messages/Rules*, which ask the LLM to produce responses that adhere to certain rules, avoid specific problematic behavior, or encourage productive behavior (specify preferred personality) [43].
   b) **Adding Examples:** If the LLM is being used to solve the same task repeatedly, we can add examples of solved problems to the input context. This is traditionally referred to as a prompting technique - if no examples are added, it is called zero-shot prompting[6], if one example is added, it is called one-shot prompting, and if multiple examples are added, it is called few-shot prompting [44].
   c) **Tool/API Knowledge:** If the LLM is being used to solve a task that requires tool-use, we can add documentation and/or examples of tool-use to the input context [10].
2) **Adaptive Augmentation:** Adding knowledge or information to the input context based on the LLM's context (or just the last message). This can be used for adding knowledge, adding examples, adding memory, and providing knowledge about tool use.

a) **Adding Knowledge:** Information can be retrieved from a knowledge base and added to the input context [8]. LLMs can also generate their search queries based on the input context [45], or simply generate the search results (generate the knowledge) [46].
b) **Adding Examples:** If the LLM is being used to solve many tasks, we can retrieve examples of the appropriate task from a knowledge base and add them to the input context [42], [47]. Examples can also be generated by other LLMs [48], [49].
c) **Adding Memory:** As LLMs have limited context length, we can create an external memory store that can be accessed to provide additional historical information to input prompts [50]. Complex LLM memory sub-systems have also been explored [51], [52].
d) **Tool/API Knowledge:** LLMs can be used to create tools (for example, executable scripts), which can then be provided for use to other LLMs, or for its own use [53]. Depending on one's perspective, all Adaptive Augmentation can be considered to be a kind of tool-use (a tool for searching, retrieving/setting memory, etc.).

2) *LLM Output Augmentation:*

While LLM Input Augmentation is useful for mitigating hallucinations, LLM Output Augmentation is used in conjunction with tools, to either give the LLM agency to influence its environment or to offload computations that LLMs are unsuitable for, such as math. When giving LLMs agency via tools, it is important to pass back feedback on whether the tool was used correctly, whether the intended effect was achieved, or to plan future tool-use [54], [55]. In R. Yamauchi *et al.* [56] and Zhibin Gou *et al.* [11], the authors ask the LLM to output python code to verify its work. The Python code is executed externally, and then the output is passed back to the model. This way, the model can verify its work and make any changes necessary.

Describing all this as "Output Augmentation" is a bit troublesome, because the LLM only "sees" this added information when generating its next message, while this "output" is in its input context. However, we still make this distinction as the reason for augmenting - off-loading computations or getting feedback from the environment, is meaningfully different from reasons for use in LLM input augmentation.

### III.B. *Prompting Techniques*

Prompting can be thought of as the act of creating an input prompt for an LLM. In the context of task-oriented LLM systems, ***prompt engineering*** can be defined as iteratively creating and adapting a prompt for a given LLM and task pair.

The way an LLM is prompted significantly affects task performance [31], [32]. There are many surprising results in this area, such as letting an LLM know that solving a task "is very important to my career" can improve task performance [57].

Such results can be explained by research such as R. Hendel *et al.* [58], which shows that in-context learning creates task vectors or representations within the LLM that increase the probability of correct task completion. Other research has shown that it is possible to "search" for prompts that are more likely to lead to success, analogous to finding task vectors that are more likely to lead to success. In A. Zou *et al.* [59], the authors were able to procedurally find adversarial prefixes, which when augmented to prompts result in LLMs breaking their alignment and engaging in unsafe behavior.

All of these are examples of modifying the prompt without changing the actual task/problem definition, to make the successful completion of the intended task more likely. However, researchers are prone to modifying prompts in a manner that changes the task, instead of modifying prompts in a manner that improves task performance.

For example, when using Chain-of-Thought prompting [60][7], or when asking LLMs to think step-by-step [6] - the task meaningfully changes. It goes from instructing LLMs to *give me an answer now* to asking it to first plan out a solution, and then share an answer. This is a different task being solved, even though the final deliverable (the answer) is the same. It should be a given that LLMs have different capabilities for different tasks.

This is not to say that we shouldn't instead solve equivalent tasks that LLMs are more suitable to, but that it is problematic to have prompt modification (that leaves instructions/task definition intact) to instruction modification in the same category. Thus, we make the distinction between prompt engineering and instruction engineering -

1. ***Prompt Engineering***: Prompt engineering is the act of modifying the prompt without changing

---

[6]This can also sometimes refer to a prompt that changes the instruction to make it more suitable for LLMs, such as Chain-of-thought [6]

[7]As described in the paper, one is to also provide in-context examples, but this is unnecessary, as shown in T. Kojima *et al.* [6]

the actual task/problem definition or adding relevant knowledge/information, to make the successful completion of the intended task more likely. We restrict the addition of relevant knowledge/information to LLM augmentation to avoid an overlap.

2. *Instruction Engineering*: Instruction engineering is the act of modifying the prompt in a manner that changes the task/problem to an equivalent task/problem that the LLMs are more suitable for, such that the final deliverable (the answer) is the same.

In M. Besta *et al.* [61], the authors describe a taxonomy of techniques for improving reliability in task-oriented text generation –

1. **Input-Output**: The LLM is directly being instructed to respond with the result for a prompted task.
2. **Input-Output with additional steps**: The LLM is being instructed to perform additional steps before or after generating a result for a prompted task, like reflecting on its response and refining it, or creating a plan [62], [60].
3. **Single Input-Many Output**:[8] The LLM is passed the same input prompt multiple times, and the most consistent answer (similar to voting) is chosen as the final answer. [63].
4. **Input with Non-linear intermediary steps**:[9] The LLM branches into multiple paths (via variations of an input prompt), generating multiple responses as additional steps, and then merges them into a single response. [15], [16].
5. **Tree of Thoughts**: A complex method described in S. Yao *et al.* [64], where many intermediate *thought branches* are explored, backtracked, and pruned until a final answer is settled on.
6. **Graph of Thoughts**: A complex method described in M. Besta *et al.* [61], where *intermediate thoughts* are modeled as a connected graph, and the LLM traverses the graph to settle on a final answer.

Later, in the Discussion (§ IV), we introduce the idea of Linear and Non-Linear contexts for prompting techniques (§ IV.B), show how it can lead to an Agent-centric Projection of any Prompting Technique (§ IV.C), and discuss its implications.

---

[8]Referred to as *Multiple CoTs* in M. Besta *et al.* [61]
[9]This is not described in M. Besta *et al.* [61]

## III.C. *Uncertainty Estimation for Task Performance*

**Uncertainty Estimation** can be thought of as the task of predicting how "correct" the output from an LLM within a task-oriented LLM system is, for a given input prompt and LLM pair. The output from an LLM is a function of its training data, the input prompt, and the sampling/decoding procedure. So, to determine the uncertainty of an LLM generation, we should focus on these three factors. The sampling/decoding procedure can be standardized and thus ignored.

Both the training data and the input prompt are natural language and thus not easy to work with. The training data poses even more challenges due to its sheer size. LLMs have shown to be very capable of tackling these challenges, and thus it may be viable to train or finetune an LLM to predict uncertainty. This would then pair every LLM in a task-oriented LLM system with a sister model tasked with predicting uncertainty, and the system would adapt as needed. It is not clear how exactly this sister model would be trained, or how it would output uncertainty.

Current research in this can be organized into 3 categories: consistency-based estimation, machine learning estimation, and LLM generated estimation.

1. *Consistency-based Uncertainty Estimation*: This approach involves estimating uncertainty based on the consistency of the outputs generated by the model. It can be achieved either through multiple samplings of the model's responses or by analyzing the probability distribution across the vocabulary tokens used in the generation process.

    In S. Mo and M. Xin [65], the authors use uncertainty as a factor to guide exploration through a tree of thoughts (or possible paths to arriving at a solution). Uncertainty was computed by applying Monte Carlo Dropout during inference, sampling multiple times, and measuring the variance in generations. Higher variances were considered to have higher uncertainty.

2. *Machine Learning-based Uncertainty Estimation*: This approach involves training a machine learning model to predict uncertainty. The input features for this model can be the hidden layer activations of the LLM, or the output logits of the LLM, or both. The output of this model can be a single value (a scalar) or a vector of values (a profile).

    In H. Y. Fu *et al.* [66], the authors propose the use of normalized probabilities reported by the LLM

across the label space for Closed-set generation tasks as a "confidence" score [66, Equation (1)]. For open-ended generation, they use negative-log likelihood (NLL) as a measure of confidence [66, Equation (2)]. These confidence scores were then used to train a *meta-model* that predicts a *confidence profile* for unseen tasks, which was found to be effective with room for improvement.

In A. Azaria and T. Mitchell [67], the authors also train an analogous *meta-model* - a feedforward neural network that takes several hidden layer activations from a large language model as input and outputs a true/false classification. True and false here refer to whether the input prompt contains a factual statement. This method, termed SAMPLA, was found to be effective at predicting the truthfulness of factual statements, thus implying that the hidden layer activations of LLMs "know" if they're engaging in uncertain generations.

3. *Verbalized Uncertainty Estimation*: In this method, Large Language Models (LLMs) are directly tasked with expressing or indicating their own level of uncertainty in their responses. This involves the model generating language that explicitly conveys how certain or uncertain it is about the information it is providing. This can be combined with machine learning approaches (by fine-tuning the LLM to predict uncertainty) or with consistency-based approaches to create a hybrid approach.

   In S. Lin *et al.* [68], the authors use consistency (from output logits) as a heuristic to create a labeled dataset of responses to math problems, and associated uncertainty. They then try finetuning an LLM to predict uncertainty and show that this method shows promise.

   In M. Xiong *et al.* [69], the authors find that a hybrid approach that combines consistency-based uncertainty estimation and verbalized uncertainty estimation is more effective than either alone.

In an ideal world, a task-oriented LLM system can predict the certainty with which it can complete a task. If task success is uncertain, the system may refuse the task, ask for help, or adapt itself to improve certainty. If task-oriented LLM systems are to be deployed in the real world, or we want multiple such systems to collaborate, this is a necessary capability. If we incorrectly predict uncertainty, we may either end up with false positives (where the system refuses to complete a task it could have completed) or false negatives (where the system attempts to complete a task it cannot complete). The current research in this area explores uncertainty estimation for simpler tasks and does not offer us a clear path to achieving uncertainty estimation for complex tasks.

## IV. Discussion

In this section, we discuss the implications of our findings, and what they suggest for future research in task-oriented LLM systems. We begin by discussing **Evaluation** (§ IV.A) in task-oriented LLM systems. While our findings help us look at the breadth of the parameters that can be tweaked when designing task-oriented LLM systems unless we can effectively evaluate each across reliability and efficiency, we wouldn't know how to traverse this design space.

Next, we discuss and define **Linear and Non-linear context** (§ IV.B) in the context of LLM systems, and how this can allow us to create an **Agent-Centric Projection of Prompting Techniques** (§ IV.C). We share examples and discuss the implications of this projection, and conjecture cross-pollination of research findings in the area of prompting and multi-agent collaboration. Finally, we discuss the implications of all this for **Synthetic Training Data Generation** (§ IV.D), about how current research in prompting can readily be used to generate synthetic training data for task-oriented LLM systems.

### IV.A. *Evaluating LLM Systems*

Current research on evaluation focuses on evaluating LLMs themselves. There is research on the evaluation of knowledge and capabilities, alignment, safety, etc. [39]. Current research on prompting or augmenting task-oriented LLMs focuses primarily on accuracy, often the percentage of tasks completed successfully, though most at least comment on computational/cost efficiency. [38], [64], [70], [65], [71].

This is because, in task-oriented LLM systems, task success is not a given, and must be measured. In Computer Science, where algorithms are required to produce correct outputs, task success is a given, and algorithms are instead evaluated on computational efficiency. However, as we progress to deploying task-oriented LLM systems in the real world, we also need to start evaluating the efficiency of our design choices in such systems.

In LLM systems, techniques and systems that need to generate fewer tokens to complete a task can be considered to be more computationally efficient. In M. Besta *et al.* [61], they use *thought volume* as a metric, which indicates the total amount of intermediary *thoughts* generated before completing the task. This is a step in the right direction; however, they assume that all *thoughts* are equally computationally expensive, which is not true. Using total tokens generated to complete a task, as described in our thought experiment, should be a better metric. It should also be noted that token vocabulary differs across different LLMs, and thus, the total tokens generated should be normalized to a common vocabulary.

As demonstrated in G.-I. Yu *et al.* [72], batching inference requests and other algorithmic improvements can yield significant generation throughput increases on the same hardware and power budget. Keeping this in mind, it could be that a system that generates more total tokens to complete a task but uses less energy overall, through differing algorithmic choices. It is therefore important to combine the total tokens generated with the total energy used to complete a task, to create a more meaningful metric.

Also, to evaluate task-oriented LLM systems, it is important to have a common interface between the task and the system. As we saw in Prompting (§ III.B), and LLM Augmentation (§ III.A), what is passed to the LLM when instructing it to solve a task, significantly influences the result. It then becomes important to standardize the interface between the task or problem definition and the LLM system, to evaluate each system fairly. It then becomes the responsibility of the LLM system to ensure it chooses the right LLM, prompting technique, augmentations, etc. to solve the task. The focus should be on evaluating the *system*.

## IV.B. *Linear and Non-linear Context in LLM Systems*

In Section II.A, we defined a minimal task-oriented LLM system to include a context store (see Figure 2). The context store includes a sequence of messages $M = \{(C_n, R_n)\}_{n=1}^{N}$ such that the first message $R_1$ is generated using only the initial context $C_1$ and later each response $R_n$ in the sequence is generated considering all previous context-response pairs $\{(C_1, R_1), (C_2, R_2), ..., (C_{n-1}, R_{n-1})\}$.

Using this definition, we can now classify every prompting technique for reliable, task-oriented text generation into two categories –

1. **Prompting techniques with Linear context** – where there is only one possible continuous sequence of messages $M = \{(C_n, R_n)\}_{n=1}^{N}$ that contains all generated messages and input contexts in the correct chronological order.

    All **Input-Output** and **Input-Output with additional steps** techniques, (as described in § III.B) can be classified as having a linear context, as they all involve a single continuous sequence of messages.

    For example, consider a simplified version of Self-Refine, first described in A. Madaan *et al.* [62]. Here, the initial context $C_1$ contains the instruction for the task, and the response $R_1$ is iteratively refined until a stop condition is met. The refinement involves all previous context-response pairs $\{(C_1, R_1), (C_2, R_2), ..., (C_{n-1}, R_{n-1})\}$, such that the last response $R_n$ is generated using all previous context-response pairs.

2. **Prompting techniques with Non-linear context** – where there cannot always be one continuous sequence of messages that contains all input context and generated messages in the correct chronological order. Instead, there can be multiple branches of conversation possible, each with its own continuous sequence of messages $\{M_1, M_2, ..., M_n\}$.

    All **Single Input-Many Output**, **Input with Non-linear intermediary steps**, **Tree of Thoughts**, and **Graph of Thoughts** techniques, (as described in § III.B) can be classified as having a non-linear context, as they all potentially involve sequences of conversation $M$.

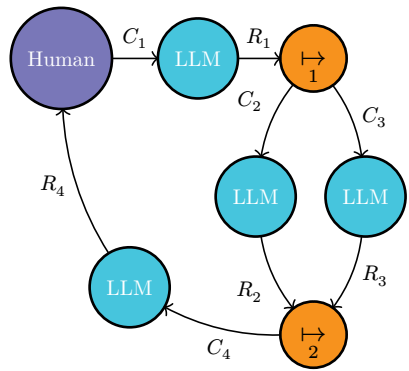

Figure 4: An example of a prompting technique with non-linear context.

For example, consider a simplified version of BRANCH-SOLVE-MERGE, first described in S. Saha *et al.* [15], and as visualized in Figure 4. The figure depicts a task-oriented LLM system that helps the user (the human) make decisions. First, the human first instructs the system to make a decision. The system uses the instructions to create an input context for an LLM (context $C_1$) and uses it to generate a response $R_1$. $R_1$ is then used to create two new prompts (via an algorithmic transformation depicted in the figure as $\mapsto_1$), one where the LLM is tasked to reflect on the drawbacks of this decision (in context $C_2$) and another where the LLM is tasked with reflecting on the benefits of this decision (in context $C_3$). Finally, another prompt is created where both reflections (responses $R_2$ and $R_3$) are considered (using another algorithmic transformation $\mapsto_2$) to create a new prompt (context $C_4$) which is used to generate a final decision inside response $R_4$. $R_4$ is then reported to the user as the final decision.

As long as any task-oriented system is using LLMs, it will always have one or more continuous streams of messages $M$ as described. This means all task-oriented LLM systems and all prompting techniques can be classified as having either linear or non-linear contexts.

This classification may have some meaningful consequences, as described in the next section.

### *IV.C. Agent-Centric Projection of Prompting Techniques*

In the previous section (§ IV.B), we classify all prompting techniques and all resulting task-oriented LLM systems they bore into either having a linear or non-linear context. This decision and the overall definition have the following implications -

1. Research on techniques for reliable, task-oriented text generation that involve sequential context can be modeled as a kind of two-agent system (human instructing the LLM being the second agent, as we also see in [13], [38]).

2. Research on techniques for reliable, task-oriented text generation that involves non-sequential context can be modeled to be a kind of Multi-agent system, where each "branch" of conversation $M$ can be considered to have occurred with a different agent.

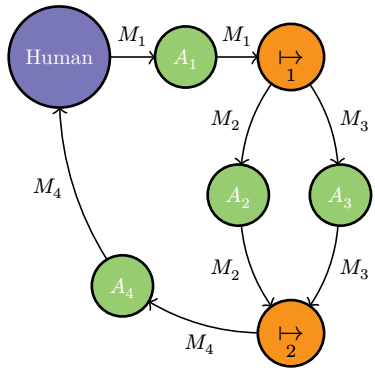

Figure 5: The prompting technique from Figure 4 is modeled as a multi-agent system.

For example, In Figure 5, we show how the prompting technique from Figure 4 can be modeled as a multi-agent system. Each continuous linear sequence of messages $M_1 = \{C_1, R_1\}$, $M_2 = \{C_2, R_2\}$, $M_3 = \{C_3, R_3\}$, and $M_4 = \{C_4, R_4\}$ can be considered to have occurred with a different agent. This way, we can model or view every prompting technique with non-linear context as a multi-agent system.

It would also help to note, that each continuous sequence of messages $M_n$ in Figure 5 is essentially a minimal task-oriented LLM system, as described in Section II.A. This means that we can substitute each such minimal system with a more complicated task-oriented system if needed.

In Figure 6, we show a more realistic example of a multi-agent system, designed to replicate the behavior of the prompting technique from Figure 4. Here, the major changes are that the agents communicate with each other using tools, meaning all communication is bidirectional (say, if an agent wants to ask a clarifying question) and that the algorithmic transformations $\mapsto_1$ and $\mapsto_2$ are now present each as a tool available to Agents $A_1$ and $A_4$ respectively. This system might behave exactly like the system in Figure 4 most of the time but may prove to be more resilient to unexpected circumstances as each component is more "intelligent".

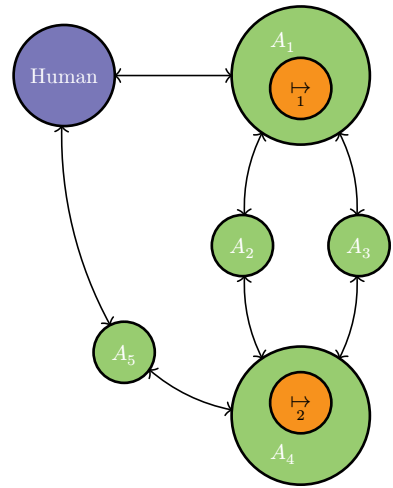

Figure 6: A more realistic projection of the prompting technique from Figure 4 as a multi-agent system.

As all prompting techniques can be projected to such multi-agent systems, we can conjecture that –

> **Conjecture 4**
>
> Results from prompting techniques involving non-sequential context, can predict similar results from multi-agent systems designed to replicate the same behavior.

This projection or view allows us to generalize all such techniques, and apply learnings from one technique to another, and even learnings from multi-agent systems. For example, if new LLM-based multi-agent collaboration research shows that "A Process Supervising Agent is all you Need", then we can immediately apply that result to the prompting technique described in BRANCH-SOLVE-MERGE from S. Saha *et al.* [15].

This can apply the other way around as well: research on a new evaluation metric that focuses on computational and/or energy efficiency as described in Section IV.A for multi-agent systems can now be applied to compare prompting techniques instead.

But it does not end there – because of how flexible natural language is, all non-sequential contexts can also be projected to sequential contexts. For example, say four agents engage in adversarial interaction as described in [13, § 4.2.2], where they argue about a decision until they reach a consensus. The benefit of this interaction paradigm is that each agent can be instructed to look at the problem from diverse perspectives.

This interaction can be elicited within a sequential context, where the LLM is prompted with the same decision-making problem, but with additional instructions to share a turn-by-turn dialogue where four individuals argue about the decision until they reach a consensus. This has been demonstrated in Zhenhailong Wang *et al.* [73], where a single LLM instance is prompted to produce a transcript of multiple personas (agents) interacting with each other to solve a task. The authors call this "self-collaboration".

> **Conjecture 5**
>
> Any result that utilizes multi-agent systems can predict similar results using prompting techniques (such as self-collaboration) designed to replicate the multi-agent interaction pattern within a sequential context.

However, as discussed in our thought experiment, this requires a single LLM to process a large amount of context which could instead be delegated to multiple LLM-based agents, which may not be ideal for non-trivial tasks. This may imply that multi-agent systems should be preferred in such cases.

*IV.D.* ***Implications for Synthetic Training Data***

This equivalence of prompting techniques with linear and non-linear context (and multi-agent systems) can have profound implications when you consider that all LLMs are trained on "sequential context", i.e., trained on a continuous sequence of text –

> **Conjecture 6**
>
> Synthetically generated "self-collaboration" traces or transcripts from successful attempts at solving tasks using prompting techniques involving non-sequential context or multi-agent collaboration, is high-quality training data for LLMs, especially for downstream use in multi-LLM agent systems and with prompting techniques involving non-linear context.

If this is found to be true, we could immediately start generating high-quality synthetic data based on existing prompting technique research and use it to train LLMs. This could be a significant advancement in the field of task-oriented LLM systems.

This idea can be taken further - Imagine if we can take the requirements of a completed software project on GitHub, along with pull requests/issue commentary, commit messages, commit diffs in chronological order, and perhaps use LLMs to fabricate communication between collaborators - wouldn't the resulting manuscript, perhaps made to resemble a theatre play script, be effective training data?

> **Conjecture 7**
>
> Taking existing problems and associated real-world deliverables (intermediate and final) and interpolating interaction artifacts between collaborators as a transcript or trace can create high-quality synthetic training data, specifically for downstream use in multi-LLM agent systems.

## V. Conclusion

In this section, we discuss the **limitations** (§ V.A) of this survey, the **key insights** (§ V.B) we gained, and the **implications for future research** in this area (§ V.C), ending with a summary of our work.

*V.A.* ***Limitations***

This scoping survey sheds light on the possibilities and potential in the design of task-oriented Large Language Model systems. The survey, while informative, is not exhaustive. For the current draft, the scope is limited to specific design parameters of task-oriented LLM systems. The thought experiment, which, though insightful, may contain assumptions that could affect its validity. For example, the thought experiment underemphasizes critical aspects such as task decomposition and planning capacities of LLMs, without which the number of collaborating agents is irrelevant.

Furthermore, the current draft has limitations in thoroughly addressing the breadth of existing research on multi-agent interaction and collaboration paradigms. This gap also extends to a lack of comprehensive discussion on the ethical and societal considerations inherent in the deployment of task-oriented LLM systems. Additionally, while the survey conjectures and comments on areas concerning training methodologies, training data, and alignment, it does not consider these topics in depth, leaving room for more detailed exploration in future work.

*V.B. Key Insights*

1. **Minimal Task-oriented LLM Systems**: The concept of a minimal LLM system, inclusive of a context store, elucidates the fundamental architecture needed for effective task-oriented AI systems. This foundation is crucial in ensuring results from architecturally diverse, advanced LLM systems can be applied to others.

2. **Extensive tool-use and Collaboration**: *Extensive tool-use* is the capacity to predict outcomes in the environment over many instances of planned tool-use, where the influence of the tool *may* be asynchronous and out-of-order. If LLM-based agents use tools to communicate with each other, then extensive tool-use may be a requirement of extensive collaboration. *Extensive collaboration* refers to a large number of LLMs-based agents collaborating, in a manner that may require some agents to ascribe and track unobservable states of multiple other agents.

3. **Prompt Engineering and Instruction Engineering**: Clearly differentiating tuning the prompt without altering the actual task or problem definition (prompt engineering) and modifying the task to an equivalent task[10] more suitable for LLM systems (instruction engineering) is essential to precise communication and understanding of research in this area.

4. **Linear and Non-linear Context in LLM Systems**: Prompting techniques and resulting task-oriented LLM systems can be classified into having either linear or non-linear context.

5. **Agent-centric Projection of Prompting Techniques**: There is a lot of research on prompting techniques that have non-sequential context, which can be modeled or viewed as a multi-agent system. This view may help us better generalize and evaluate results in the domain of task-oriented LLM systems.

6. **Evaluating LLM Systems**: Emphasizing metrics for evaluation such as computational efficiency (perhaps via total tokens generated) and energy consumption in evaluating these systems and techniques is a step towards mapping the design space, in turn enabling practical applications.

    Focus should be on evaluating the *system* by standardizing the interface between the task/problem definition and the system, so the system can adapt (choose the LLM, prompting technique/augmentation) to the given task.

7. **Uncertainty Estimation for Task Performance**: The research into uncertainty estimation is an essential advancement towards developing self-adaptive task-oriented LLM systems.

*V.C. Implications for Future Research*

1. **Cross-Pollination in Prompting and LLM-based Multi-agent Systems**: The agent-centric projection of prompting techniques may allow us to cross-pollinate research findings in these areas.

2. **Multi-Agent Collaboration**: Our findings point towards a need for more research in multi-agent LLM collaborations. Understanding how these agents can work together efficiently and effectively is key to solving more complex, real-world tasks.

3. **Synthetic Training Data Generation**: The idea of creating synthetic training data by simulating agent interactions or from traces of researched prompting techniques is intriguing. This approach could provide a rich source of data for training more advanced LLMs, particularly for multi-agent environments.

4. **Real-world Applications and Ethical Considerations**: As these systems become more capable, their deployment in real-world scenarios becomes more feasible. With this comes the need for rigorous ethical considerations, especially concerning autonomy, decision-making, and human-AI interaction.

In summary, this survey has led us to three primary categories of conjectures that may be pivotal in advancing our understanding and application of task-oriented Large Language Models (LLMs). Firstly, we conjecture the critical role of multi-agent collaboration in enhancing the capabilities of LLM systems. This involves exploring how less ca-pable, tool-augmented language models can, through collaboration, surpass the performance of more individually capable models; allowing us to achieve productive task-oriented LLM systems without significant breakthroughs in LLM training and architecture. Secondly, we emphasize the potential of various prompting techniques, especially in non-sequential contexts, and their implications for multi-agent system design. This includes the exploration of how results from non-sequential prompting can inform and predict the outcomes in multi-agent systems. We also explore the possibilities of synthetic training data generation, positing that such data, particularly those derived from traces of successful task-solving attempts using advanced prompting techniques or multi-agent interactions, can serve as a valuable resource for training more sophisticated LLMs. These conjectures collectively highlight the dynamic interplay between collaboration, prompting strategy, and training data in shaping the future of task-oriented LLM systems. Finally, we emphasize the importance of the task of uncertainty estimation in task-oriented LLM systems, evaluate the current state of research, and find that we are still far from a solution that would allow us to create self-adaptive task-oriented LLM systems.

---

[10]one with the same final deliverable